\documentstyle[prl,aps,psfig,epsfig,multicol]{revtex}
\begin{document}
\def\be{\begin{equation}}
\def\ee{\end{equation}}
\def\bc{\begin{center}}
\def\ec{\end{center}}
\def\bea{\begin{eqnarray}}
\def\eea{\end{eqnarray}}
\newcommand{\bleq}{\ifpreprintsty
                   \else\bibitem{BAH}
A.-L. Barab\`asi, R. Albert and H. Jeong, {\it Physica} {\bf
272A}, 173 (1999).
                   \end{multicols}\vspace*{-3.5ex}\widetext{\tiny
                   \noindent\begin{tabular}[t]{c|}
                   \parbox{0.493\hsize}{~} \\ \hline \end{tabular}}
                   \fi}
\newcommand{\eleq}{\ifpreprintsty
                   \else
                   {\tiny\hspace*{\fill}\begin{tabular}[t]{|c}\hline
                    \parbox{0.49\hsize}{~} \\
                    \end{tabular}}\vspace*{-2.5ex}\begin{multicols}{2}
                    \narrowtext \fi}

\title{ Number of loops of  size $h$ in growing scale-free networks.}
\author{Ginestra Bianconi and Andrea Capocci}
\address{Institut the Physique Th\'eorique, Universit\'e de Fribourg
P\'erolles,CH-1700, Switzerland}
\maketitle
\begin{abstract}
The hierarchical structure of scale-free  networks has been investigated focusing on the scaling 
of the number $N_h(t)$ of loops of size $h$ as a function of the system size.
In particular we have found the analytic expression for the scaling of $N_h(t)$  in the 
Barab\'asi-Albert scale-free network. We have performed numerical simulations on the scaling 
law for $N_h(t)$ in the BA network and in other growing scale free networks, such as the bosonic 
network and the aging nodes network.
We show that in the bosonic network and in the aging node network the phase transitions in the 
topology of the network are accompained by a change in the scaling of the number of loops with the 
system size.
\end{abstract}
\begin{multicols}{2}
\narrowtext


Natural and social systems ranging from protein-protein interactions \cite{Naturepp} 
to the World Wide Web \cite{BA} can be represented as {\em scale-free} (SF) networks 
\cite{RMP,Strogatz,DoroRev,New_rev}, that is, as a set of nodes connected by links with
special statistical properties. 
These systems are characterized by large fluctuations in the individual {\em degrees}, 
i.e. in the number of links pointing or leaving each node.
In a SF network, the statistical distribution $P(k)$ of the degree $k$ is a power law 
$P(k)\sim k^{-\gamma}$. Accordingly, if $\gamma<3$, the individual degree 
distribution displays infinite variance in the limit of infinite network size.

Besides this striking property, models traditionally defined on regular lattices display 
peculiar effects when defined on a scale-free topology instead \cite{Cohen12,Vespignani,Stauffer,Doro_ising,Leone,Ising}.
Such phenomena have increased the interest in finding universal mechanisms to
generate SF networks, which seem to arise independently in so diverse contexts.

Recently, these properties have been reproduced by the Barab\'asi-Albert (BA) 
dynamical model of a random network. Its simple and natural algorithm, based on a ``preferential
attachment'' rule $\cite{BA,BAH}$, triggered an avalanche of research activity in the 
field and generated a rich ``zoology'' of SF network models sharing the same fundamental 
ingredients. Though recently alternative mechanisms \cite{Caldarelli,Static} for the generation of scale-free 
network have been proposed, in this work we restrict our attention to growing scale-free networks 
with ``preferential attachment''.
But the degree distribution is not the only relevant topological quantity. For the characterization of 
networks it is necessary to look at the motifs\cite{Milo,Klemm,Szabo} recurrent on it and in particular at the number 
of loops of length larger then three \cite{Vespignani_loops}.
In this work we derive a general formula for the number of loops of size $h$ 
that generalizes the known result concerning triangular loops \cite{circles} ($h=3$) 
for a BA network. In particular, we find that the number of loops $N_h(t)$ of size $h$ at time $t$ in a BA 
network scales as $N_h(t)\sim (m/2\log(t))^\psi$.
In addition, we show that our formula is robust when tested on other growing 
SF network models, such as the bosonic network model (BN) $\cite{bose}$ and the aging nodes 
model (AN) $\cite{aging}$. In the bosonic network the scaling of the number of loops with the system 
size changes below the Bose-Einstein phase transition becoming a power-law scaling of the type $N_h(t)\sim t^{\xi}$.

The BA model \cite{BA,BAH} was the first and simplest algorithm generating scale-free 
undirected networks. In this model, a new node is added to the network at each time step, and it is connected by 
a fixed number of links $m$ to highly connected existing nodes (preferential attachment). 
According to this rule, the probability $\Pi_i$ that an existing node $i$ at time $t$ acquires 
a new link is assumed to be proportional to its degree $k_i(t)$. This reads
\be
\Pi_i = m \frac{k_i(t)}{\sum_j k_j(t)}.
\label{pref}
\ee
The model can be analyzed by a mean field approximation \cite{BA}. 
By this approach, one finds that the average degree of a node $i$ that entered the 
network at time $t_i$ increases with time as 
a power-law
\be
k_i(t)=\frac{m}{2}\sqrt{\frac{t}{t_i}}.
\label{ki}
\ee
A network built in this way displays a power law degree distribution $P(k) \sim k^{-3}$.
In order to investigate the inhomogeneous topology of the network, we define 
a loop of size $h$ (a $h$-loop) as a closed path of $h$ links that visits
each intermediate node only once.

In the case $m>2$, the BA scale-free network is a very compact network, with
loops of any size.
As the network evolves, new loops are introduced in the network. 
By definition, new loops include the newly added node: indeed, a new $h$-loop 
is formed if the new node is connected to two nodes already connected 
by a self-avoiding path of size $h-2$.
We indicate with $p_{i,k}$ the probability that the nodes
${i,k}$, attached to the network  at time ${t_i, t_k}$, are
connected by a link. The rate at which new loops of length $h$ are
formed is given by the probability that the new node $k$ is linked 
to two existing nodes $i$ and $j$ times the probability $P_{i,j}^{h-2}(t)$ 
that they are already connected by a self-avoiding path of size $h-2$.
Therefore, we write the following rate equation for the average number of
$h$-loops $N_h(t)$ 
\be
\frac{\partial <N_h(t)>}{\partial t}=\frac{1}{2} \sum_{i=1}^N
\sum_{j=1}^N p_{i,k} p_{j,k} P_{ij}^{h-2}(t) \label{Nhdyn}, 
\ee
where the factor $\frac{1}{2}$ takes into account that 
each pair of nodes $i$, $j$ has been counted twice in the sums.

On the other hand, two nodes belong to a $h$-loop if there is a link
between them and, besides it, they are connected by a self-avoiding 
path of length $h-1$. Let $P^{h-1}_{i,j}(t)$ be the probability that this path exists;
thus, the probability that the link between node $i$ and $j$ belongs to a 
$h$-loop is given by $p_{i,j} P^{h-1}_{i,j}(t)$. 
We obtain the average number $<N_h(t)>$ of $h$-loops in the system
times $2h$ by summing this quantity over all the nodes $i$, $j$ in the network. 
In fact, each loop has been counted $2h$ times, because
there are $h$ nodes in the loop, and two possible directions.
Hence, we can write
\be
<N_h(t)>=\frac{1}{2 h}\sum_{i=1}^N \sum_{j=1}^N p_{i,j}
P_{ij}^{h-1}(t). \label{Nh-1} 
\ee
By replacing Eq. $(\ref{ki})$ in Eq. ($\ref{pref}$), the 
probability that the node $i$ is attached to node $k$ is
\be
p_{i,k}= \frac{m}{2}\frac{1}{\sqrt{t_i t_k}}.  \label{pik} \ee  
Moreover, the probability that a node $k$ is
connected with nodes $i$ and $j$ is proportional to the
probability that $i$ and $j$ are already connected, i.e.
\be
p_{i,k}p_{j,k}= m\frac{1}{2 t_k} p_{i,j}. \label{p2p1} 
\ee
By replacing this result in $(\ref{Nhdyn})$ and by the
definition $(\ref{Nh-1})$, we obtain
 \be
\frac{\partial<N_h(t)>}{\partial t} =\frac{m}{2 t}
(h-1)<N_{h-1}(t)>. \label{rec} 
\ee 
Consequently, the rate at which new loops of size $h$ are introduced 
in the system is proportional to the mean number of loops of size $h-1$. 

Equation $(\ref{rec})$ has a recursive structure that allows its integration
without any detailed information about the probabilities $P_{i,j}^{h}(t)$.
In fact, the rate at which new loops of length $h$ are formed can be expressed only 
in terms of the number of loops of minimal size (i.e. $h=3$),
\be
\frac{\partial^{h-3} <N_h(\zeta)>}{\partial \zeta^{h-3}}=(h-1)!
<N_{3}(\zeta)> \label{nxi}
\ee 
with $\zeta=\frac{m}{2} \log(t)$.  
The number of triangular loops\  $<N_{3}(\zeta)>$ can be computed directly, 
for the triangular loops increase in time following
 $(\ref{Nhdyn})$,
\bea 
\frac{\partial<N_3(t)>}{\partial t}
&=&\frac{1}{2}\sum_{i=1}^N \sum_{j=1}^N p_{i,k}p_{j,k}p_{i,j}.
\eea
Since $p_{i,j}$ is given by eq. $(\ref{pik})$, approximating sums by integrals we write
the rate equation in the form
\bea
\frac{\partial<N_3(t)>}{\partial t}&=&\frac{1}{2}\left(\frac{m}{2}\right)^3\int_0^t dt_i
\int_0^t dt_j \frac{1}{t_i}\frac{1}{t_j}\frac{1}{t} \nonumber
\\&=& \frac{1}{2}\left(\frac{m}{2}\right)^3 \frac{1}{t}
[\log(t)]^2. \label{n3dyn}
\eea 
Integrating $(\ref{n3dyn})$ we find, in agreement with $\cite{circles}$,
\be
<N_3(t)>=\frac{1}{3!} \left[\frac{m}{2}\log(t)\right]^3.
\label{n3a}
\ee
\begin{figure}
\centerline{\epsfxsize=2.5in \epsfbox{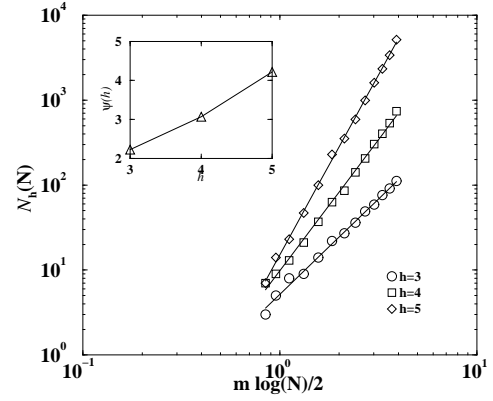}} \caption{Scaling of $N_h(t)$ as a function  
in a BA network for $t$ up to $t=10^4$. The Inset shows the values of the measured exponent 
$\psi$ defined in$(\ref{psi_1})$ for $h=3,4,5$. } \label{loops_ba.fig}
\end{figure}

Using eq. $(\ref{n3a})$ in eq. $(\ref{nxi})$, we compute the number of loops of size $h$, $<N_h(t)>$ 
and we find 
\bea \label{Nh_t}
<N_h(t)>&=&\left[\frac{m}{2}\log(t)\right]^h
(1+O(\zeta^{-1})).\nonumber
\\ &\sim &\left[\frac{m}{2}\log(t)\right]^h. 
\eea

The expression for the scaling of $N_h(t)$ with the system size $t$ in a BA network 
does not suggest a practical way to measure $N_h(t)$.
To this purpose, one has to study the symmetrical adjacency matrix $a$ of the network,
whose generic element $a_{ij}$ is defined by $a_{ij} =1$ if $i$ and $j$ are connected and 
$a_{ij}=0$ otherwise.
Knowing this matrix, one directly measures the number of paths starting from a node $i$ and returning 
on it after $h$ steps that visit intermediate nodes only once.
According to this argument, the term $N_h(t)$ has a dominating term of the type
$\sum_i (a^h)_{i,i}/(2h)$ and subdominant terms excluding all trivial contributions
coming from paths intersecting on themselves.
Let us assume that the network does not contain self-loops, i.e. $a_{ii}=0$ for all $i$ of the 
network. In this case, for $h=3$ we simply have
\be
N_3=\frac{1}{6}\sum_i (a^3)_{ii}.
\label{n3}
\ee 
For  $h=4$ and $h=5$, by simple arguments it is possible to show that
\be
N_4=\frac{1}{8}\left[\sum_i (a^4)_{ii}-2\sum_i(a^2)_{ii} (a^2)_{ii}+\sum_i (a^2)_{ii}\right]
\label{n4}
\ee
and that
\be
N_5=\frac{1}{10}\left[\sum_i (a^5)_{ii}-5\sum_i(a^2)_{ii} (a^3)_{ii}+5\sum_i (a^3)_{ii}\right].
\label{n5}
\ee
Using relations $(\ref{n3})$, $(\ref{n4})$, $(\ref{n5})$, we can directly measure $N_h(t)$ 
for $h=3,4,5$ in the BA scale-free network model to check our analytical results.

In Fig. $\ref{loops_ba.fig}$ we show that the scaling of $N_h(t)$ for a BA  network with $m=2$ follows 
\be
N_h(t) \sim \left(\frac{m}{2}\log(t)\right)^{\psi(h)}.
\label{psi_1}
\ee
The measured effective exponent $\psi(h)$ reported in the Inset of Fig. $\ref{loops_ba.fig}$  grows 
with $h$ as expected. Nevertheless, $\psi(h)$ differs from the predicted asymptotic behavior $\psi(h)=h$ 
in a significant way. This can be explained by the fact that we are considering networks of size up 
to $t=10^4$ nodes and we are still far form the asymptotic behavior $t\rightarrow \infty$.

To check the robustness of the scaling relations of $N_h(t)$,
we have measured the number of loops of size $h=3,4,5$ in two alternative growing 
SF network models: the bosonic network (BN)  $\cite{bose}$ and the aging nodes network (AN)  $\cite{aging}$. 

In the BN bosonic network, each node $i$ is assigned an innate quality, represented by a
random 'energy' $\epsilon_i$ drawn form the probability distribution $p(\epsilon_i)$.
The attractiveness of each node $i$ is then determined jointly by its connectivity $k_i$ and 
its energy $\epsilon_i$. In particular, the probability that node $i$ acquires a link at time $t$ 
is given by
\be
\Pi_i=\frac{e^{-\beta\epsilon_i}k_i(t)}{\sum_je^{-\beta\epsilon_j}k_j(t)},
\ee 
i.e. low energy, high degree nodes are more likely to acquire new links.
The parameter $\beta=1/T$ in $ \Pi_i$ tunes the relevance of the quality with respect 
to the degree in the acquisition probability of new links.
Indeed, for $T\rightarrow \infty$ the probability $\Pi_i$ does not depend any more on the energy 
$\epsilon_i$ and the BN model reduces to the BA model. 
On the other hand, in the limit $T \rightarrow 0$ only the lowest energy node has non zero 
probability to acquire new links.
Moreover, in Ref.$\cite{bose}$ it has been shown that the connectivity distribution in this network 
model can be mapped on the occupation number in a Bose gas. According to this analogy, one would 
expect a corresponding phase transition in the topology of the network at some temperature value $T_c$.

In fact, for energy distributions such that ($p(\epsilon)\rightarrow 0$ for $\epsilon \rightarrow 0$),
one observes a critical temperature $T_c$. 
For $T>T_c$ the system is in the ``fit-get-rich''(FGR) phase, where nodes with lower energy acquire 
links at a higher rate that higher energy nodes, while for $T<T_c$ a ``Bose-Einstein condensate''(BE) 
or ``winner-takes-all'' phase emerges, where a  single nodes grabs a finite fraction of all the links.
We simulated this model assuming   
\bea
p(\epsilon)=(\theta+1)\epsilon^\theta\ \  & \mbox{and} &\ \  \epsilon\in (0,1) 
 \label{p_epsilon}
\eea
where $\theta=1$. 
For this distribution, it has been shown that $T_c(\theta=1)=0.7$ \cite{bose}.
In particular, in Fig. $\ref{loop_cond.fig}$ we report $N_h(t)$ for $h=3,4,5$ as a function of 
the number of nodes $t$ in the network, at temperature $T=1.5$ and $T=0.5$ respectively.

According to simulations, in correspondence with the phase transition, the scaling 
of $N_h(t)$ drastically switches from a BA-like behavior
\bea
N_h(t)\propto\left(\frac{m}{2}\log(t)\right)^{\psi(h)}\ & \mbox{for}&\  T>T_c 
\eea
to a power-law  scaling
\bea
N_h(t)\propto t^{\xi(h)}\ &\mbox{for}&\  T<T_c.
\eea
\begin{figure}
\centerline{\epsfxsize=2.5in \epsfbox{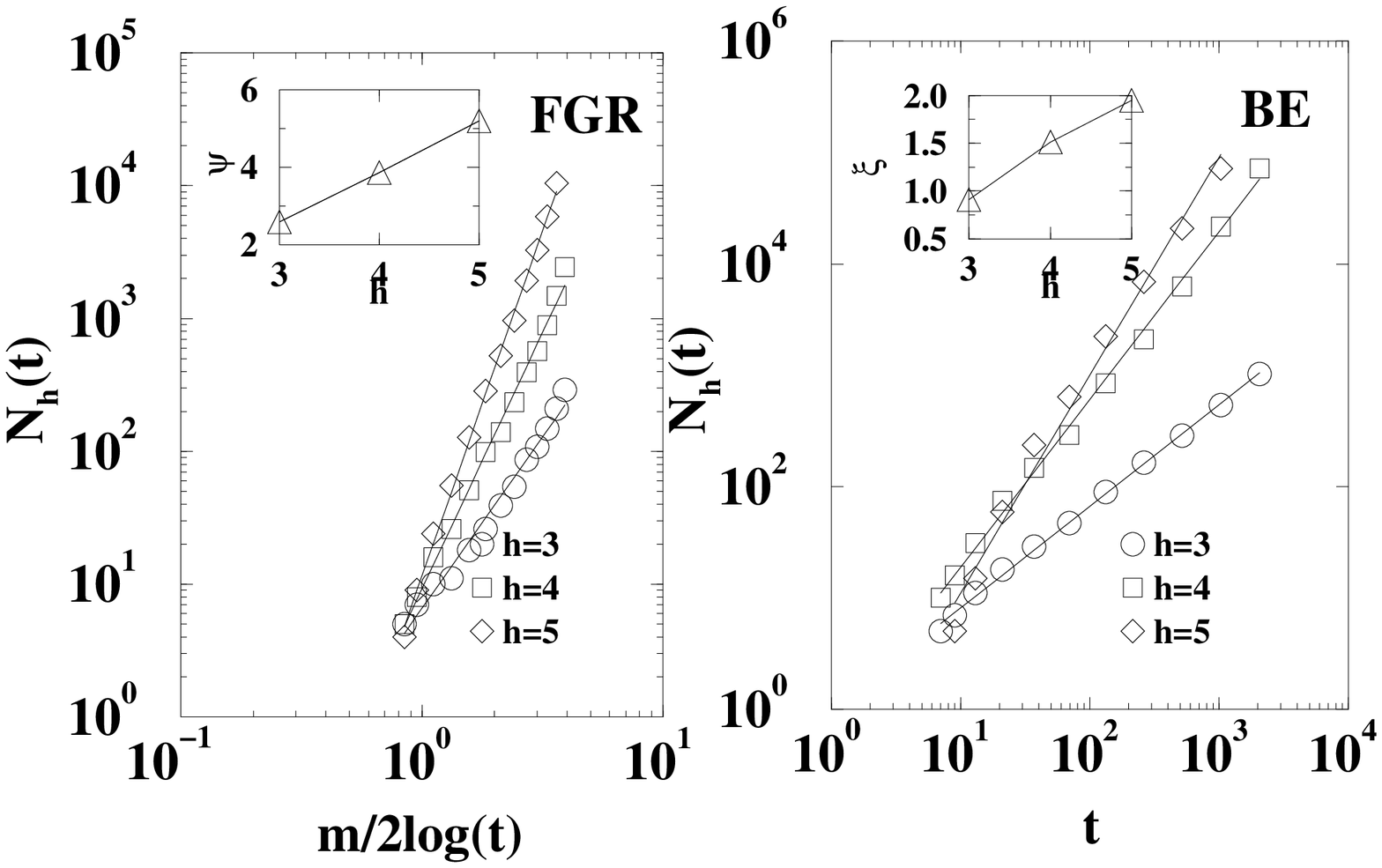}} \caption{Scaling of the number of loops $N_h(t)$ of size $h$ with the system size $t$ in the FGR and in the BE phase of a bosonic network. The plotted data have been obtained  for a network with $p(\epsilon)=2\epsilon$ and $\epsilon\in (0,1)$ (which has $T_c=0.7$ [19]) at  $T=1.5$ (FGR phase) and at $T=0.5$ (BE phase). In the Insets we report the value of the exponents $\psi$ (2.6,3.9,5.2)  and $\xi$ (0.78,1.28,156) found in the simulations for $h=3,4,5$ respectively.} \label{loop_cond.fig}
\end{figure}

We have measured the scaling of $N_h(t)$ with the system size $t$ also for a growing network with 
aging nodes (AN) introduced in $\cite{aging}$.
The model has been motivated by the observation that in many real networks, e.g. the scientific citations 
network, old nodes are less cited than recent ones.
In the goal of representing this feature, the probability $\Pi_i$ to attach a link to 
a node $i$ arrived in the network at time $t_i$ is modified to be
\be
\Pi_i=\frac{(t-t_i)^{-\alpha}k_i(t)}{\sum_j (t-t_j)^{-\alpha}k_j(t)}.
\ee  
where $\alpha$ is an external parameter. As in the BA model, a new node is connected to
$m$ existing nodes.
The resulting structure of such a network strongly depends on the constant $\alpha$.
For $\alpha<1$, the degree distribution is a power law $P(k) \sim k^{-\gamma}$ with an exponent 
monotonically increasing from  $\gamma=2$ in the limit $\alpha \rightarrow -\infty$ to $\gamma\rightarrow \infty$ 
in the limit  $\alpha \rightarrow 1$: on the other hand, for $\alpha>1$ no power law is observed in the
 degree distribution. Therefore, this model reproduces a SF network only in the region $\alpha>1$.

Moreover, as observed  in $\cite{aging}$, in the limit $\alpha\rightarrow -\infty$ the oldest node 
is connected to an increasing fraction of all the links, reminding the condensation observed in a bosonic 
network.
To take into account this phenomenon, we introduce a value $\alpha^*$ such that for $\alpha<\alpha^*$ 
the fraction of links attached to the most connected node exceeds a finite threshold $F$.
We expect the scaling of $N_h(t)$ to be different in the three regions
$\alpha>1$, $\alpha\in(\alpha^*,1)$ and $\alpha<\alpha^*$.
\begin{figure}
\centerline{\epsfxsize=2.5in \epsfbox{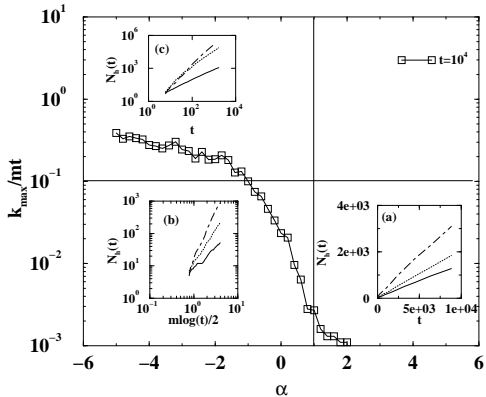}} \caption{The order parameter $k_{max}/mt$ for a network with aging of the nodes, size $t=10^4$ and $m=2$. We distinguish between three region of the phase space:$\alpha>1$, $\alpha\in(-1,1)$ and $\alpha<-1$. In the Insets we report the typical behavior of  $N_h(t)$ as a function of $t$ for $h=3,4,5$ in the three regions. } \label{aging.fig}
\end{figure}
We measured the total fraction of links $k_{max}/mt$ attached to the oldest node in a network with $m=2$ 
and $t=10^4$ nodes. The threshold has been fixed at $F=0.1$, in order to distinguish the 'condensate' 
phase from the simple scale-free phase. The value for $\alpha^*$ was found to be $\alpha^*=-1$.
We then measured the number of loops of size $h=3,4,5$ for networks made of up to $10^4$ nodes in 
the three ranges of value of $\alpha$.
We have observed that, for $\alpha>1$, the number of loops of size $h$ scales linearly with $t$ (at
least for $h=3,4,5$). In the Inset (a) of Fig.~$\ref{aging.fig}$ we report the data for $\alpha=1.5$.
On the contrary, for $\alpha \in (-\alpha^*,1)$ we measured the scaling
\be
N_h(t)\propto \left(\frac{m}{2}\log(t)\right)^{\psi(h)}
\ee
with $\psi(h)$ a monotonic function of $h$. In Inset (b) of Fig.~$ \ref{aging.fig}$, data for $\alpha=0.5$
are reported.
Finally, in the region $\alpha<\alpha^*=-1$, $N_h(t)$ becomes proportional to a power-law 
of the system size, as it is shown in Inset (c) of Fig.~$ \ref{aging.fig}$, referring to the case 
$\alpha=-5$.

In conclusion, we have introduced the number of $h$-loops $N_h(t)$ for a 
network of $t$ nodes as a characterizing quantity for random networks.
We have observed that in the BA scale-free networks $N_h(t)$ scales as a power law of the logarithm 
of the system size.
Moreover, we have observed that indeed this scaling seems to be a marking feature of growing scale-free 
networks with preferential attachment. In particular, topological phase transitions in the bosonic network 
and in network including aging nodes are accompanied by a drastic change in the scaling of $N_h(t)$ with the 
system size.

The authors are grateful to Yi-Cheng Zhang and Alessandro Vespignani for their useful comments.
This work has been partially supported by the European Commission - Fet Open project COSIN
IST-2001-33555.


\end{multicols}
\end{document}